\documentclass[sigconf]{acmart}
\usepackage{booktabs} 
\usepackage{amsmath,url,graphicx,amssymb}
\usepackage{amsfonts}
\usepackage{ctable}
\usepackage{multirow}
\usepackage{algorithm}
\usepackage{algpseudocode}
\usepackage{pifont}
\usepackage{color}
\usepackage{subfigure}

\newcommand{\mylistbegin}{
  \begin{list}{$\bullet$}
   {
     \setlength{\itemsep}{-2pt}
     \setlength{\leftmargin}{1em}
     \setlength{\labelwidth}{1em}
     \setlength{\labelsep}{0.5em} } }
\newcommand{\mylistend}{
   \end{list}  }

\newcommand{\eg}{\textit{e.g.}}

\newcommand{\ie}{\textit{i.e.}}

\newcommand{\wrt}{\textit{w.r.t.~}}

\copyrightyear{2018} 
\acmYear{2018} 
\setcopyright{acmcopyright}
\acmConference[KDD 2018]{24th ACM SIGKDD International Conference on Knowledge Discovery \& Data Mining}{August 19--23, 2018}{London, United Kingdom}
\acmBooktitle{KDD 2018: 24th ACM SIGKDD International Conference on Knowledge Discovery \& Data Mining, August 19--23, 2018, London, United Kingdom}
\acmPrice{15.00}
\acmDOI{10.1145/3219819.3219821}
\acmISBN{978-1-4503-5552-0/18/08} 

\begin{document}
\title[\textit{ClusChurn}]{I Know You'll Be Back: Interpretable New User Clustering and Churn Prediction on a Mobile Social Application}
\author{Carl Yang\footnotemark[1]{\small \footnotemark[2]}, Xiaolin Shi{\small \footnotemark[2]}, Jie Luo{\small \footnotemark[2]}, Jiawei Han\footnotemark[1]}
       \affiliation{
       \institution{\footnotemark[1]University of Illinois, Urbana Champaign, 201 N Goodwin Ave, Urbana, IL 61801, USA}
       \institution{{\small \footnotemark[2]}Snap Inc., 64 Market St, Venice, CA 90291, USA}
       \institution{\footnotemark[1]\{jiyang3, hanj\}@illinois.edu, {\small \footnotemark[2]}\{xiaolin.shi, roger.luo\}@snap.com}
       }

\setlength{\floatsep}{4pt plus 4pt minus 1pt}
\setlength{\textfloatsep}{4pt plus 2pt minus 2pt}
\setlength{\intextsep}{4pt plus 2pt minus 2pt}
\setlength{\dbltextfloatsep}{3pt plus 2pt minus 1pt}
\setlength{\dblfloatsep}{3pt plus 2pt minus 1pt}
\setlength{\abovecaptionskip}{3pt}
\setlength{\belowcaptionskip}{2pt}
\setlength{\abovedisplayskip}{2pt plus 1pt minus 1pt}
\setlength{\belowdisplayskip}{2pt plus 1pt minus 1pt}
\renewcommand{\shortauthors}{Carl Yang, Xiaolin Shi, Jie Luo and Jiawei Han}
\settopmatter{printacmref=false, printfolios=false}

\begin{abstract}
As online platforms are striving to get more users, a critical challenge is user churn, which is especially concerning for new users. In this paper, by taking the anonymous large-scale real-world data from Snapchat as an example, we develop \textit{ClusChurn}, a systematic two-step framework for interpretable new user clustering and churn prediction, based on the intuition that proper user clustering can help understand and predict user churn. Therefore, \textit{ClusChurn} firstly groups new users into interpretable typical clusters, based on their activities on the platform and ego-network structures. Then we design a novel deep learning pipeline based on LSTM and attention to accurately predict user churn with very limited initial behavior data, by leveraging the correlations among users' multi-dimensional activities and the underlying user types. \textit{ClusChurn} is also able to predict user types, which enables rapid reactions to different types of user churn. Extensive data analysis and experiments show that \textit{ClusChurn} provides valuable insight into user behaviors, and achieves state-of-the-art churn prediction performance. 
The whole framework is deployed as a data analysis pipeline, delivering real-time data analysis and prediction results to multiple relevant teams for business intelligence uses. 
It is also general enough to be readily adopted by any online systems with user behavior data.
\end{abstract}


\maketitle
{\small \textbf{ACM Reference Format:}\\
Carl Yang, Xiaolin Shi, Jie Luo, Jiawei Han. 2018. I Know You'll Be Back: Interpretable New User Clustering and Churn Prediction on a Mobile Social Application. In \textit{KDD 2018: 24th ACM SIGKDD International Conference on Knowledge Discovery \& Data Mining, August 19--23, 2018, London, United Kingdom.} ACM, New York, NY, USA, 9 pages. https://doi.org/10.1145/3219819.3219821}
\section{Introduction}
\label{sec:intro}
Promoted by the widespread usage of internet and mobile devices, hundreds of online systems are being developed every year, ranging from general platforms like social media and e-commerce websites to vertical services including news, movie and place recommenders.
As the market is overgrowing, the competition is severe too, with every platform striving to attract and keep more users. 

While many of the world's best researchers and engineers are working on smarter advertisements to expand businesses by acquisition \cite{harris2006method, moriarty2014advertising}, retention has received less attention, especially from the research community. The fact is, however, acquiring new users is often much more costly than retaining existing ones\footnote{https://www.invespcro.com/blog/customer-acquisition-retention}.
With the rapid evolution of mobile industry, the business need for better user retention is larger than ever before\footnote{http://info.localytics.com/blog/mobile-apps-whats-a-good-retention-rate}, for which, \textit{accurate}, \textit{scalable} and \textit{interpretable} churn prediction plays a pivotal role\footnote{https://wsdm-cup-2018.kkbox.events}.

Churn is defined as a user quitting the usage of a service. 
Existing studies around user churn generally take one of the two ways: data analysis and data-driven models. 
The former is usually done through user surveys, which can provide valuable insights into users' behaviors and mindsets. But the approaches require significant human efforts and are hard to scale, thus are not suitable for nowadays ubiquitous mobile apps.
The development of large-scale data-driven models has largely improved the situation, but no existing work has looked into user behavior patterns to find the reasons behind user churn. 
As a consequence, the prediction results are less interpretable, and thus cannot fundamentally solve the problem of user churn.

In this work, we take the anonymous data from Snapchat as an example to systematically study the problem of interpretable churn prediction. 
We notice that online platform users can be highly heterogeneous. For example, they may use (and leave) a social app for different reasons\footnote{http://www.businessofapps.com/data/snapchat-statistics}.
Through extensive data analysis on users' multi-dimensional temporal behaviors, we find it intuitive to capture this heterogeneity and assign users into different clusters, which can also indicate the various reasons behind their churn.
Motivated by such observations, we develop \textit{ClusChurn}, a framework that jointly models the types and churn of new users (Section \ref{sec:data}). 

To understand user types, we encounter the challenges of automatically discovering interpretable user clusters, addressing noises and outliers, and leveraging correlations among features.
As a series of treatments, we apply careful feature engineering and adopt $k$-means with Silhouette analysis \cite{rousseeuw1987silhouettes} into a three-step clustering mechanism. The results we get include six intuitive user types, each having typical patterns on both daily activities and ego-network structures. In addition, we also find these clustering results highly indicative of user churn and can be directly leveraged to generate type labels for users in an unsupervised manner (Section \ref{sec:clustering}).

To enable interpretable churn prediction, we propose to jointly learn user types and user churn. Specifically, we design a novel deep learning framework based on LSTM \cite{hochreiter1997long} and attention \cite{denil2012learning}. Each LSTM is used to model users' temporal activities, and we parallelize multiple LSTMs through attention to focus on particular user types. Extensive experiments show that our joint learning framework delivers supreme performances compared with baselines on churn prediction with limited user activity data, while it also provides valuable insights into the reasons behind user churn, which can be leveraged to fundamentally improve retention (Section \ref{sec:prediction}).

Note that, although we focus on the example of Snapchat data, our \textit{ClusChurn} framework is general and able to be easily applied to any online platform with user behavior data. A prototype implementation of \textit{ClusChurn} based on PyTorch is released on Github\footnote{https://github.com/yangji9181/ClusChurn}.

The main contributions of this work are summarized as follows:
\begin{enumerate}
\item Through real-world large-scale data analysis, we draw attention to the problem of interpretable churn prediction and propose to jointly model user types and churn.
\item We develop a general automatic new user clustering pipeline, which provides valuable insights into different user types.
\item Enabled by our clustering pipeline, we further develop a prediction pipeline to jointly predict user types and user churn and demonstrate its interpretability and supreme performance through extensive experiments.
\item We deploy \textit{ClusChurn} as an analytical pipeline to deliver real-time data analysis and prediction to multiple relevant teams within Snap Inc. It is also general enough to be easily adopted by any online systems with user behavior data.
\end{enumerate}


\section{Large-Scale Data Analysis}
\label{sec:data}
To motivate our study on user clustering and churn prediction, and gain insight into proper model design choices, we conduct an in-depth data analysis on a large real-world dataset from Snapchat.
Sensitive numbers are masked for all data analysis within this paper.

\subsection{Dataset}
We collect the anonymous behavior data of all new users who registered their accounts during the two weeks from August 1, 2017, to August 14, 2017, in a particular country. We choose this dataset because it is a relatively small and complete network, which facilitates our in-depth study on users' daily activities and interactions with the whole network. 
There are a total of 0.5M new users registered in the specific period, and we also collect the remaining about 40M existing users with a total of approximately 700M links in this country to form the whole network.

\begin{table}[h]
\small
 \centering
 \begin{tabular}{|c|c|c|}
 \hline
ID&Feat. Name&Feat. Description\\
\hline
0&chat\_received&\# textual messages received by the user\\
\hline
1&chat\_sent&\# textual messages sent by the user\\
\hline
2&snap\_received&\# snap messages received by the user\\
\hline
3&snap\_sent&\# snap messages sent by the user\\
\hline
4&story\_viewed&\# stories viewed by the user\\
\hline
5&discover\_viewed&\# discovers viewed by the user\\
\hline
6&lens\_posted&\# lenses posted to stories by the user\\
\hline
7&lens\_sent&\# lenses sent to others by the user\\
\hline
8&lens\_saved&\# lenses saved to devices by the user\\
\hline
9&lens\_swiped&\# lenses swiped in the app by the user\\
\hline
 \end{tabular}
 \caption{\label{tab:act}\textbf{\small Daily activities we collect for users on Snapchat.}}
 \vspace{-5pt}
\end{table}

We leverage two types of features associated with users, \ie, their daily activities and ego-network structures. Both types of data are collected during the two-week time window since each user's account registration. Table \ref{tab:act} provides the details of the daily activities data we collect, which are from users' interactions with some of the core functions of Snapchat: \textit{chat, snap, story, lens}. We also collect each user's complete ego-network, which are formed by her and her direct friends. The links in the networks are bi-directional friendships on the social app. For each user, we compute the following two network properties and use them as a description of her ego-network structures.
\begin{itemize}
\item Size: the number of nodes, which describes how many friends a user has.
\item Density: the number of actual links divided by the number of all possible links in the network. It describes how densely a user's friends are connected.
\end{itemize}

As a summary, given a set of $N$ users $\mathcal{U}$, for each user $u_i \in \mathcal{U}$, we collect her 10-dimensional daily activities plus 2-dimensional network properties, to form a total of 12-dimensional time series $\mathbf{A}_i$. The length of $\mathbf{A}_i$ is 14 since we collect each new user's behavioral data during the first two weeks after her account registration. Therefore, $\mathbf{A}_i$ is a matrix of $12\times14$.

\subsection{Daily Activity Analysis}
\begin{figure}[h!]
\centering
\subfigure[Daily Measures]{
\includegraphics[width=0.23\textwidth]{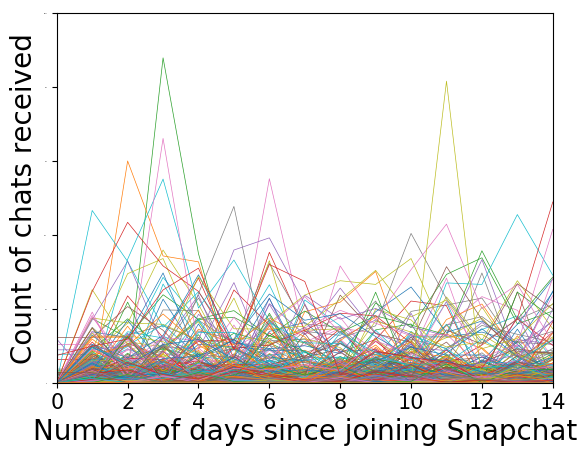}}
\subfigure[Aggregated Measures]{
\includegraphics[width=0.23\textwidth]{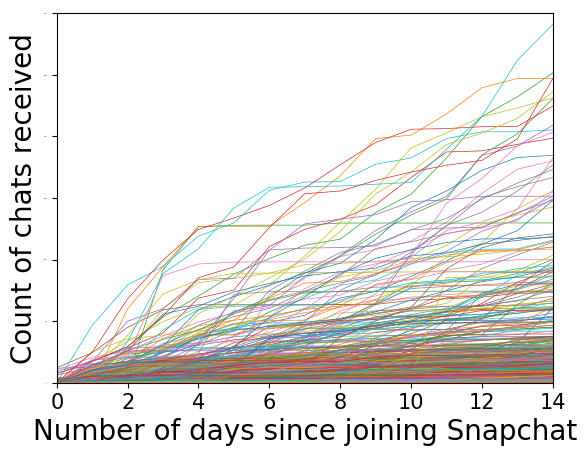}}
\vspace{-5pt}
\caption{\textbf{Activities on \textit{chat\_received} in the first two weeks. Y-axis is masked in order not to show the absolute values.}}
\vspace{-5pt}
\label{fig:act}
\end{figure}

Figure \ref{fig:act} (a) shows an example of daily measures on users' \textit{chat\_received} activities. Each curve corresponds to the number of chats received by one user every day during the first two weeks after her account registration. The curves are very noisy and bursty, which poses challenges to most time series models like HMM (Hidden Markov Models), as the critical information is hard to be automatically captured. Therefore we compute two parameters, \ie, $\mu$, the mean of daily measures to capture the activity volume, and $l$, the $lag(1)$ of daily measures to capture the activity burstiness. Both metrics are commonly used in time series analysis \cite{box2015time}. 

Figure \ref{fig:act} (b) shows the aggregated measures on users' \textit{chat\_received} activities. Every curve corresponds to the total number of chats received by one user until each day after her account registration. The curves have different steepness and inflection points. Motivated by a previous study on social network user behavior modeling \cite{ceyhan2011dynamics}, we fit a sigmoid function $y(t)=\frac{1}{1+e^{-q(t-\phi)}}$ to each curve, and use the two parameters $q$ and $\phi$ to capture the shapes of the curves. 

\begin{figure}[h!]
    \includegraphics[width=0.8\linewidth]{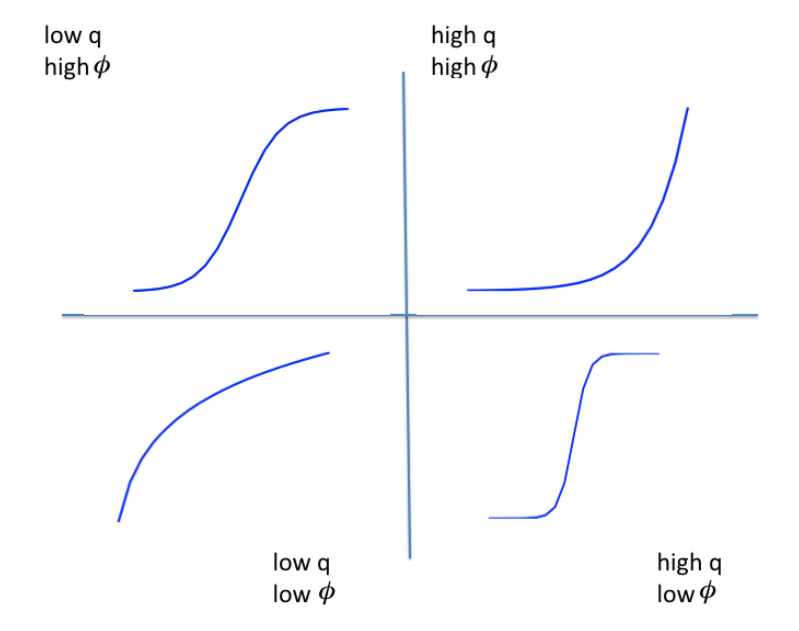}
    \vspace{-5pt}
    \caption{Main curve shapes captured by sigmoid functions with different parameter configurations.}
    \vspace{-5pt}
    \label{fig:sigmoid}
\end{figure}

Figure $\ref{fig:sigmoid}$ shows 4 example shapes of curves captured by the sigmoid function with different $q$ and $\phi$ values.After such feature engineering on the time series data, each of the 12 features is described by a vector of 4 parameters $\mathbf{f}=\{\mu, l, q, \phi\}$. We use $\mathbf{F}_i$ to denote the feature matrix of $u_i$ and $\mathbf{F}_i$ is of size $12\times4$.

\subsection{Network Structure Analysis}
In addition to daily activities, we also study how new users connect with other users. The 0.5M new users in our dataset directly make friends with a subset of a few million users in the whole network during the first two weeks since their account registration. We mask the absolute number of this group of users and use $\kappa$ to denote it.

We find these $\kappa$ users very interesting since there are about 114M links formed among them and 478M links to them. However, there are fewer than 700M links created in the whole network of the total about 40M users in the country. It leads us to believe that there must be a small group of well-connected popular users in the network, which we call the \textit{core} of a network, and in fact, this core overlaps with a lot of the $\kappa$ direct friends of new users.

\begin{figure}[h!]
\centering
\subfigure[Overlapping of core and the $\kappa$ users]{
\includegraphics[width=0.23\textwidth]{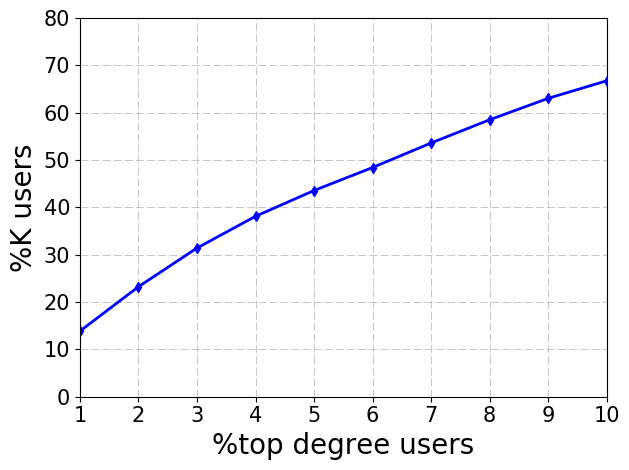}}
\subfigure[Degree distribution of the $\kappa$ users]{
\includegraphics[width=0.23\textwidth]{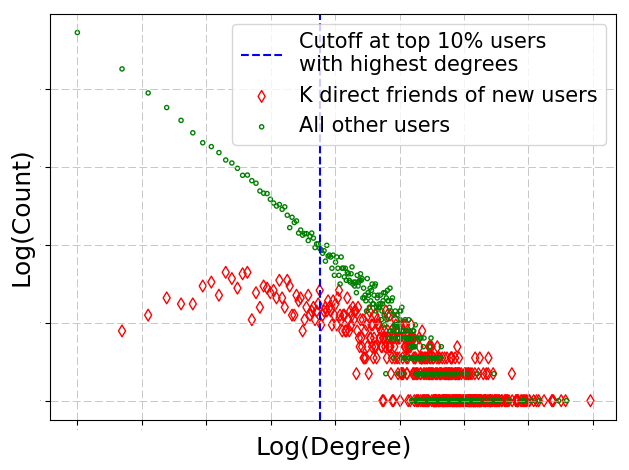}}
\vspace{-5pt}
\caption{\textbf{Most of the $\kappa$ users are within the core.}}
\vspace{-5pt}
\label{fig:core}
\end{figure}

To validate this assumption, we define the core of social networks as the set of users with the most friends, \ie, nodes with highest degrees, motivated by earlier works on social network analysis  \cite{shi2008very}. Figure \ref{fig:core} (a) shows the percentage of the $\kappa$ users within the core as we increase the size of the core from the top 1\% nodes with highest degrees to the top 10\%. Figure \ref{fig:core} (b) shows the particular degrees of the $\kappa$ users drawn together with all other nodes, ordered by degrees on the x-axis. As we can see, 44\% of the $\kappa$ users are among the top 5\% nodes with highest degrees, and 67\% of them have 10\% highest degrees. This result confirms our hypothesis that most links created by new users at the beginning of their journeys are around the network core. Since the $\kappa$ direct friends do not entirely overlap with the core, it also motivates us to study how differently new users connect to the core, and what implications such differences can have on user clustering and churn prediction.
\section{Interpretable User Clustering}
\label{sec:clustering}

In this section, we study what the typical new users are like on Snapchat and how they connect to the social network. We aim to find an interpretable clustering of new users based on their initial behaviors and evolution patterns when they interact with the various functions of a social app and other users. Moreover, we want to study the correlations between user types and user churn, so as to enable better churn prediction and personalized retention.

We also note that, besides churn prediction, interpretable user clustering is crucial for the understanding of user behaviors so as to enable various product designs, which can ultimately lead to different actions towards the essentially different types of users. Therefore, while we focus on the end task of churn prediction, the framework proposed in this work is generally useful for any downstream applications that can potentially benefit from the understanding of user types, such as user engagement promotion.

\subsection{Challenges}
Automatically finding interpretable clustering of users \wrt multi-dimensional time series data poses quite a few challenges, which makes the canonical algorithms for clustering or feature selection such as $k$-means and principal component analysis impractical \cite{han2011data}.

{\flushleft \textbf{Challenge 1: Zero-shot discovery of typical user types}}. 
As we discuss in Section \ref{sec:intro}, users are often heterogeneous. For example, some users might actively share contents, whereas others only passively consume \cite{hu2014we}; some users are social hubs that connect to many friends, while others tend to keep their networks neat and small \cite{kwak2010twitter}. However, for any arbitrary social app, is there a general and systematic framework, through which we can automatically discover the user types, without any prior knowledge about possible user types or even the proper number of clusters?

{\flushleft \textbf{Challenge 2: Handling correlated multi-dimensional behavior data}}. 
Users interact with a social app in multiple ways, usually by accessing different functions of the app as well as interacting with other users. Some activities are intuitively highly correlated, such as \textit{chat\_sent} and \textit{chat\_received}, whereas some correlations are less obvious, such as \textit{story\_viewed} and \textit{lens\_sent}. Moreover, even highly correlated activities cannot be simply regarded as the same. For example, users with more chats sent than received are quite different from users in the opposite. Therefore, what is a good way to identify and leverage the correlations among multiple dimensions of behavior data, including both functional and social activities?

{\flushleft \textbf{Challenge 3: Addressing noises and outliers}}.
User behavior data are always noisy with random activities. An active user might pause accessing the app for various hidden reasons, and a random event might cause a dormant user to be active for a period of time as well. Moreover, there are always outliers, with extremely high activities or random behavior patterns. A good clustering framework needs to be robust to various kinds of noises and outliers.

{\flushleft \textbf{Challenge 4: Producing interpretable clustering results}}.
A good clustering result is useless unless the clusters are easily interpretable. In our scenario, we want the clustering framework to provide insight into user types, which can be readily turned into actionable items to facilitate downstream applications such as fast-response and targeted user retention.

\subsection{Methods}
\begin{figure}[h!]
    \includegraphics[width=0.8\linewidth]{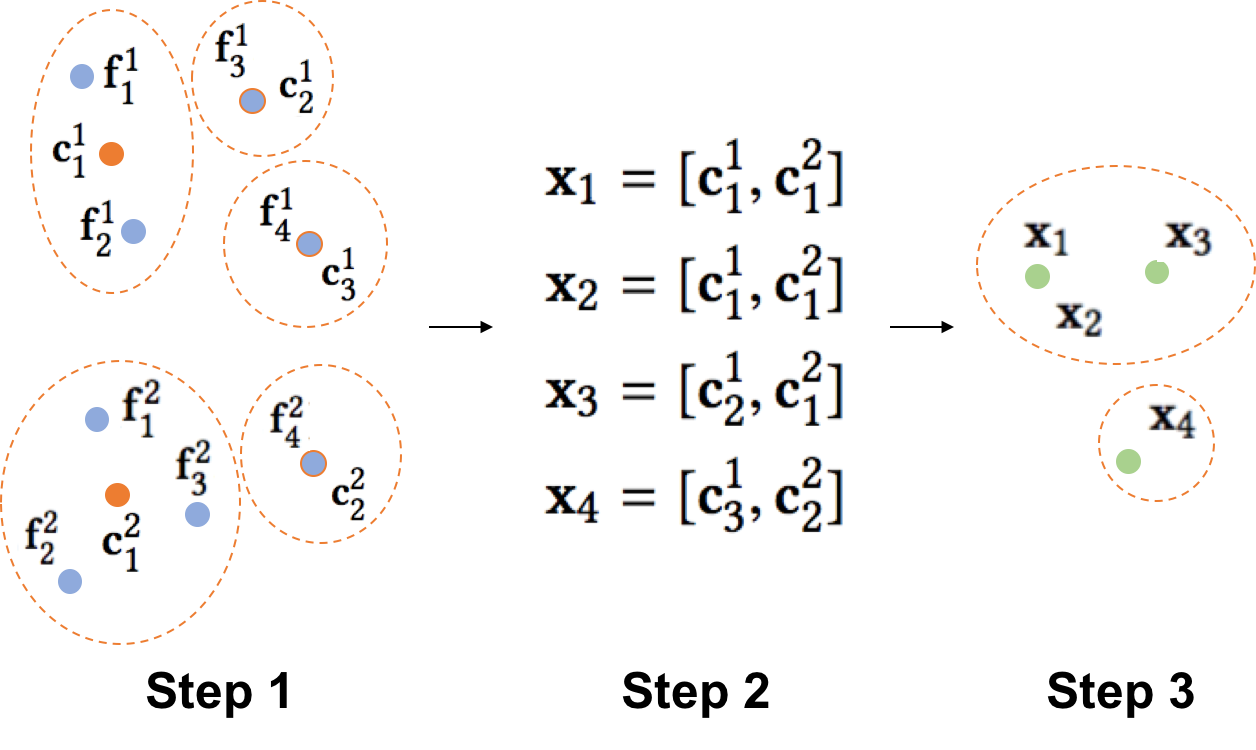}
    \caption{A toy example of our 3-step clustering framework.}
     \vspace{-15pt}
    \label{fig:clus}
\end{figure}

To deal with those challenges, we design a robust three-step clustering framework. 
Consider a total of two features, namely, $\mathbf{f}^1$ (\textit{chat\_received}) and $\mathbf{f}^2$ (\textit{chat\_sent}), for four users, $u_1$, $u_2$, $u_3$ and $u_4$. 
Figure \ref{fig:clus} illustrates a toy example of our clustering process with the details described in the following.

{\flushleft \textbf{Step 1: Single-feature clustering}}.
For each feature, we apply $k$-means with Silhouette analysis \cite{rousseeuw1987silhouettes} to automatically decide the proper number of clusters $K$ and assign data into different clusters. 
For example, as illustrated in Figure \ref{fig:clus}, for \textit{chat\_received}, we have the feature of four users $\{\mathbf{f}^1_1, \mathbf{f}^1_2, \mathbf{f}^1_3, \mathbf{f}^1_4\}$, each of which is a $4$-dimensional vector (\ie, $\mathbf{f}=\{u,l,q,\phi\})$. Assume $K$ chosen by the algorithm is 3. Then we record the cluster belongingness, \eg, $\{l^1_1=1, l^1_2=1, l^1_3=2, l^1_4=3\}$, and cluster centers $\{\mathbf{c}^1_1, \mathbf{c}^1_2, \mathbf{c}^1_3\}$. Let's also assume that for \textit{chat\_sent}, we have $K=2$, $(l^2_1=1, l^2_2=1, l^2_3=1, l^2_4=2)$ and $\{\mathbf{c}^2_1, \mathbf{c}^2_2\}$.
This process helps us to find meaningful types of users \wrt every single feature, such as users having high volumes of \textit{chat\_received} all the time versus users having growing volumes of this same activity day by day.

{\flushleft \textbf{Step 2: Feature combination}}.
We convert the features of each user into a combination of the features of her nearest cluster center in each feature. Continue our toy example in Figure \ref{fig:clus}. Since user $u_1$ belongs to the first cluster in feature \textit{chat\_received} and first cluster in feature \textit{chat\_sent}, it is replaced by $\mathbf{x}_1$, which is a concatenation of $\mathbf{c}^1_1$ and $\mathbf{c}^2_1$. $u_2$, $u_3$ and $u_4$ are treated in the same way.
This process helps us to largely reduce the influence of noises and outliers because every single feature is replaced by that of a cluster center.

{\flushleft \textbf{Step 3: Multi-feature clustering}}.
We apply $k$-means with Silhouette analysis again on the feature combinations. As for the example, the clustering is done on $\{\mathbf{x}_1, \mathbf{x}_2, \mathbf{x}_3, \mathbf{x}_4\}$. The algorithm explores all existing combinations of single-dimensional cluster centers, which record the typical values of combined features. 
Therefore, the multi-feature clustering results are the typical combinations of single-dimensional clusters, which are inherently interpretable.
 
\subsection{Results}

\begin{figure}[h!]
 \vspace{-5pt}
\centering
\subfigure[Parameter dist.]{
\includegraphics[width=0.13\textwidth]{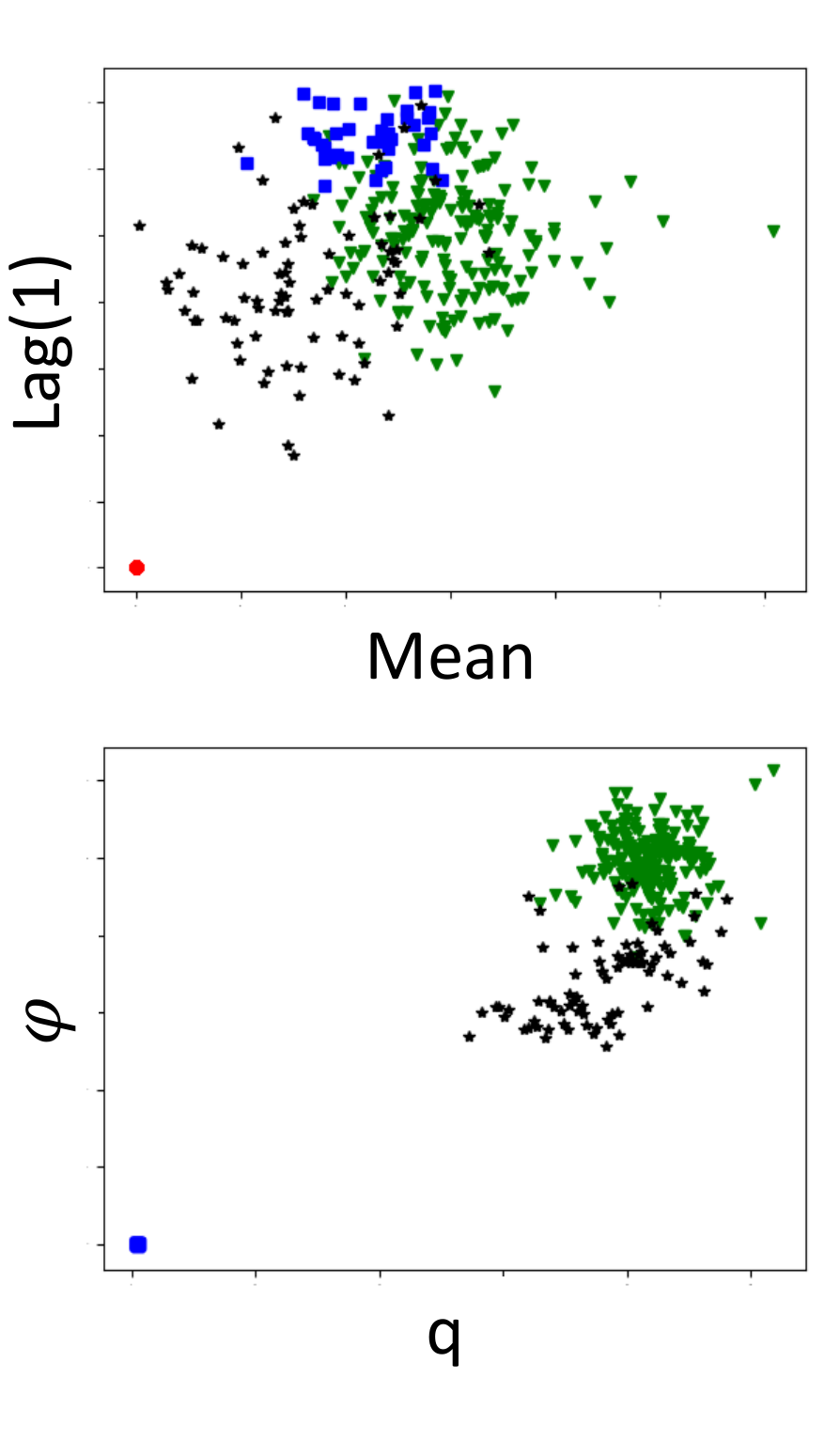}}
\subfigure[Activity patterns.]{
\includegraphics[width=0.33\textwidth]{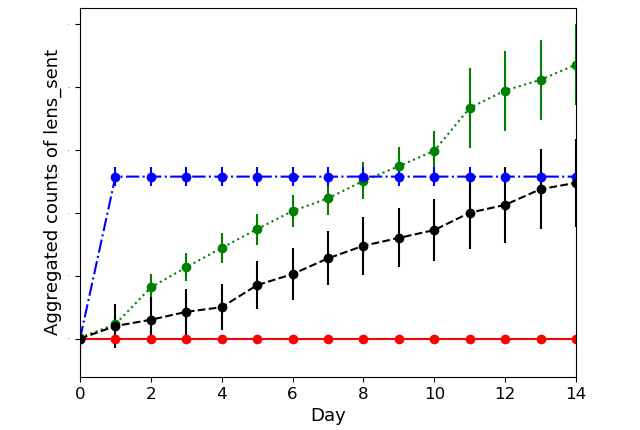}}
 \vspace{-5pt}
\caption{\textbf{4 types of users shown with different colors.}}
 \vspace{-5pt}
\label{fig:func_res}
\end{figure}

{\flushleft \textbf{Clustering on single features}}.
We first present our single-feature clustering results on each of users' 12-dimensional behaviors. Figure \ref{fig:func_res} provides the detailed results on \textit{lens\_sent} as an example. The results on other features are slightly different regarding the numbers of clusters, shapes of the curves, and numbers of users in each cluster. However, the method used is the same and they are omitted to keep the presentation concise.

Figure \ref{fig:func_res} (a) shows the four parameters we compute over the 14-day period on users' \textit{lens\_sent} activities, as they distribute into the four clusters detected by the $k$-means algorithm. The number of clusters is automatically selected with the largest average Silhouette score when $k$ is iterated from 2 to 6, which corresponds to clusters that are relatively far away from each other while having similar sizes. Figure \ref{fig:func_res} (b) shows the corresponding four types of users with different activity patterns on \textit{lens\_sent}. The first type of users (red) have no activity at all, while the second type (green) have stable activities during the two weeks. Type 3 users (blue) are only active in the beginning, and type 4 users (black) are occasionally active. These activity patterns are indeed well captured by the volume and burstiness of their daily measures, as well as the shape of the curves of their aggregated measures. Therefore, the clusters are highly interpretable. By looking at the clustered curves, we can easily understand the activity patterns of each type of users.

{\flushleft \textbf{Clustering on network properties}}.
For single-feature clustering on network properties, as we get four clusters on ego-network size and three clusters on density, there is a total of $4\times 3=12$ possible combinations of different patterns. However, when putting these two features of network properties together with the ten features of daily activities through our multi-feature clustering framework, we find that our new users only form three typical types of ego-networks. This result proves the efficacy of our algorithm since it automatically finds that only three out of the twelve combinations are essentially typical.

\begin{figure}[h!]
    \includegraphics[width=0.9\linewidth]{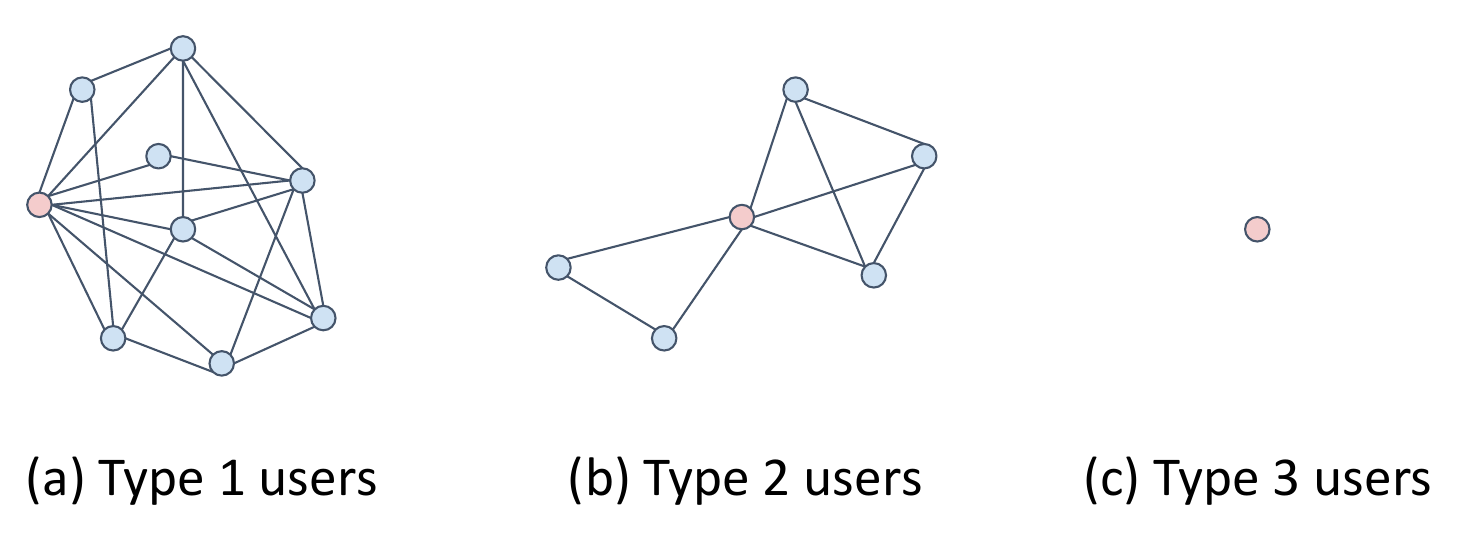}
    \caption{\small Examples of 3 types of ego-network structures.}
     \vspace{-5pt}
    \label{fig:net_res}
\end{figure}

Figure \ref{fig:net_res} illustrates three example structures. The ego-networks of type 1 users have relatively large sizes and high densities;
type 2 users have relatively small ego-network sizes and low densities; 
users of type 3 have minimal values on both measures.

\begin{figure}[h!]
    \includegraphics[width=0.7\linewidth]{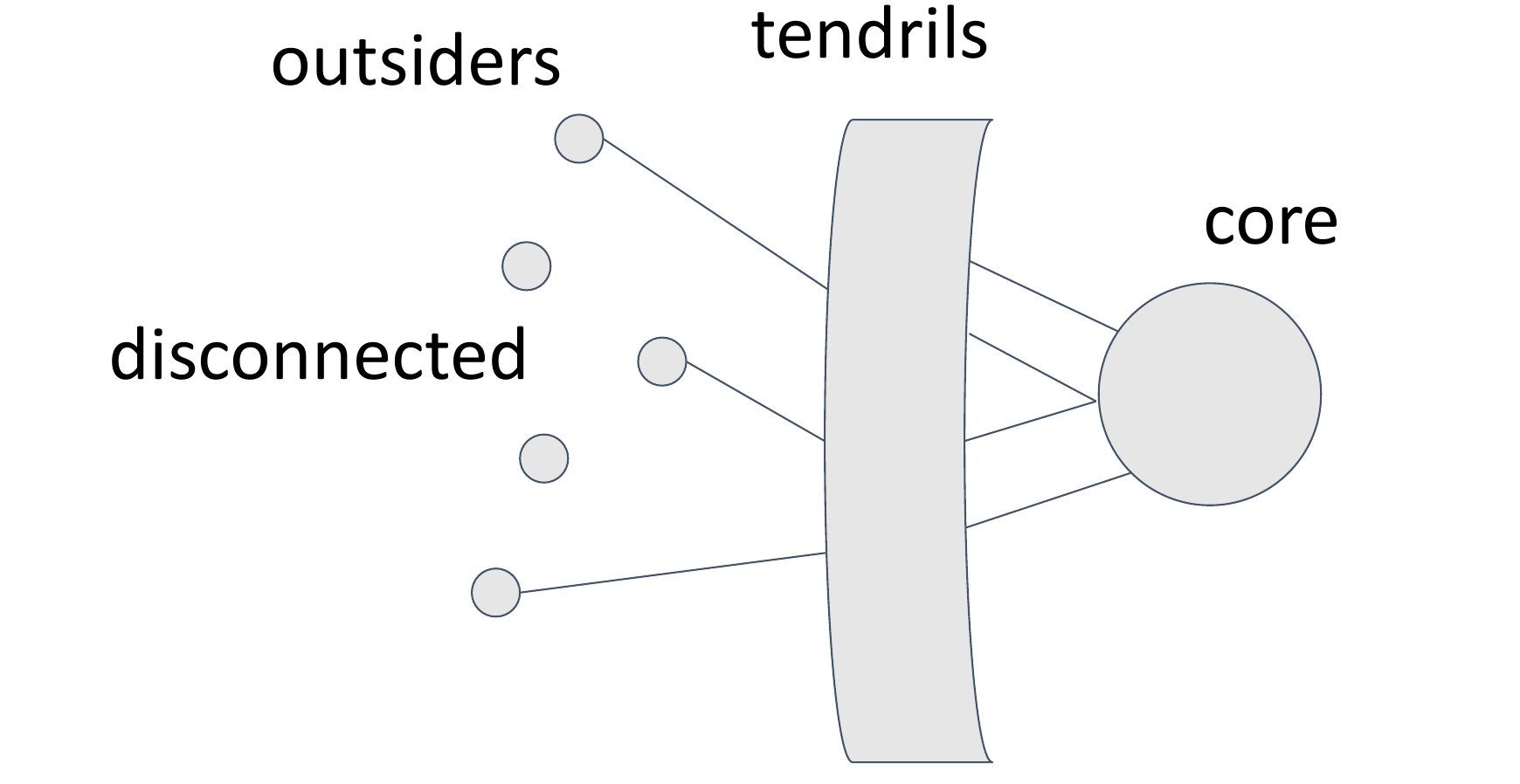}
    \caption{The whole network depicted into a jellyfish shape.}
     \vspace{-5pt}
    \label{fig:jelly}
\end{figure}

Through further analysis, we find that these three types of new users clustered by our algorithm based on the features of their ego-networks have strong correlations with their positions in the whole social network. Precisely, if we define network core as the top 5\% users that have the most friends in the entire network, and depict the whole network into a jellyfish structure as shown in Figure \ref{fig:jelly}, we can exactly pinpoint each of the three user types into the tendrils, outsiders, and disconnected parts. Specifically, type 1 users are mostly tendrils with about 58\% of direct friends in the core; type 2 users are primarily outsiders with about 20\% of direct friends in the core; type 3 users are mostly disconnected with almost no friends in the core. Such result again proves that our clustering framework can efficiently find important user types.

\begin{figure}[h!]
\centering
\subfigure[Portions]{
\includegraphics[width=0.47\linewidth]{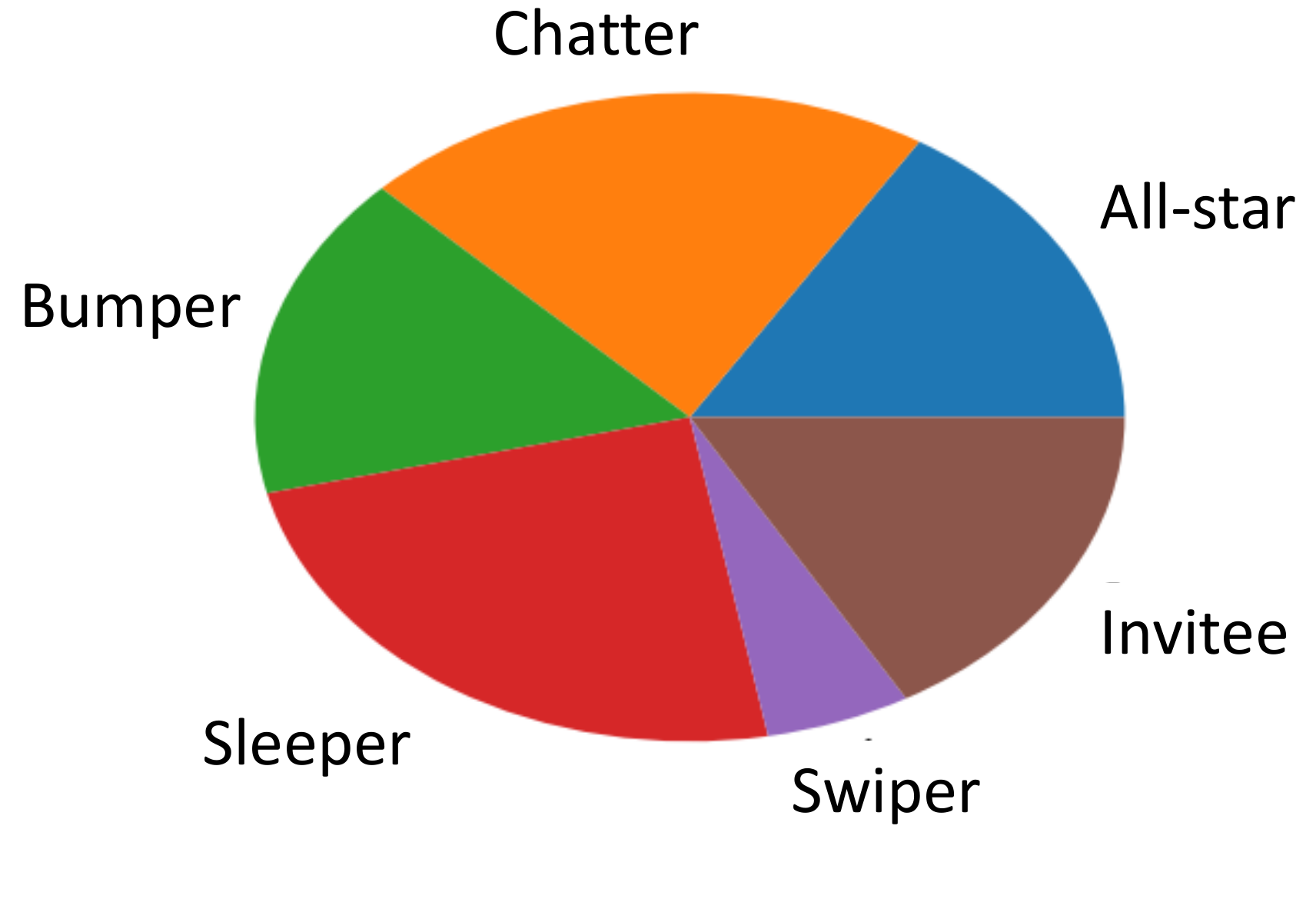}}
\subfigure[Churn rates]{
\includegraphics[width=0.47\linewidth]{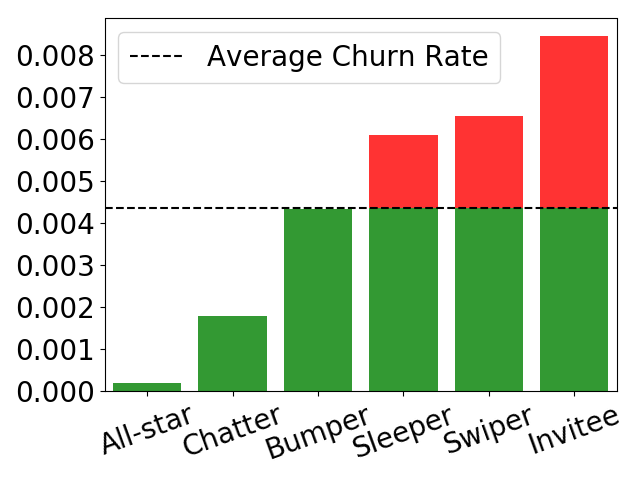}}
\caption{\textbf{Portions and churn rates of the six new user types. The y-axis is rescaled to not show the absolute values.}}
 \vspace{-5pt}
\label{fig:type}
\end{figure}

{\flushleft \textbf{Clustering on all behaviors}}.
Combining new users' network properties with their daily activities, we finally come up with six cohorts of user types, which is also automatically discovered by our algorithm without any prior knowledge. Looking into the user clusters, we find their different combinations of features quite meaningful, regarding both users' daily activities and ego-network structures. Subsequently, we are able to give the user types intuitive names, which are shown in Table \ref{tab:type}. 
Figure \ref{fig:type} (a) shows the portions of the six types of new users. 

We define a user churns if there is no activity at all in the second week after account registration.
To get more insight from the user clustering results and motivate an efficient churn prediction model, we also analyze the churn rate of each type of users and present the results in Figure \ref{fig:type} (b). The results are also very intuitive. For example, All-star users are very unlikely to churn, while Swipers and Invitees are the most likely to churn. 

\begin{table}[h]
\footnotesize
 \centering
 \begin{tabular}{|c|c|c|c|}
 \hline
ID&Type Name&Daily Activities & Ego-network Type\\
\hline
0 & All-star & Stable active chat, snap, story \& lens & Tendril\\
\hline
1 & Chatter & Stable active chat \& snap, few other acts & Tendril\\
\hline
2 & Bumper & Unstable chat \& snap, few other acts & Tendril\\
\hline
3 & Sleeper & Inactive & Disconnected\\
\hline
4 & Swiper & Active lens swipe, few other acts & Disconnected\\
\hline
5 & Invitee & Inactive & Outsider\\
\hline
 \end{tabular}
 \caption{\label{tab:type}\textbf{\small 6 types of new users and their characteristics.}}
  \vspace{-5pt}
\end{table}

Note that, our new user clustering results are highly intuitive, and in the meantime provide a lot of valuable insights. For example, the main differences between All-star users and Chatters are their activities on \textit{story} and \textit{lens}, which are the additional functions of Snapchat. Being active in using these functions indicates a much lower churn rate. The small group of Swipers is impressive too since they seem to only try out the lenses a lot without utilizing any other functions of the app, which is related to an entirely high churn rate. Quite a lot of new users seem to be invited to the app by their friends, but they are highly likely to quit if not interacting with their friends, exploring the app functions or connecting to core users. 
Insights like these are highly valuable for user modeling, growth, retention and so on.

Although we focus our study on Snapchat data in this paper, the clustering pipeline we develop is general and can be applied to any online platforms with multi-dimensional user behavior data. The code of this pipeline has also been made publicly available.


\section{Fast-Response Churn Prediction}
\label{sec:prediction}
Motivated by our user type analysis and the correlations between user types and churn, we aim to develop an efficient algorithm for interpretable new user churn prediction. 
Our analysis of real data shows that new users are most likely to churn in the very beginning of their journey, which urges us to develop an algorithm for fast-response churn prediction. The goal is to accurately predict the likelihood of churn by looking at users' very initial behaviors, while also providing insight into possible reasons behind their churn.

\subsection{Challenges}
New user churn prediction with high accuracy and limited data is challenging mainly for the following three reasons.

{\flushleft \textbf{Challenge 1: Modeling sequential behavior data.}}
As we discuss in Section \ref{sec:data}.1, we model each new user by their initial interactions with different functions of the social app as well as their friends, and we collect a 12-dimensional time series $\mathbf{A}_i$ for each new user $u_i \in \mathcal{U}$. However, unlike for user clustering where we leverage the full two-week behavior data of each user, for fast-response churn prediction, we only focus on users' very limited behavior data, \ie, from the initial few days. The data are naturally sequential with temporal dependencies and variable lengths. Moreover, the data are very noisy and bursty. These characteristics pose great challenges to traditional time series models like HMM.

{\flushleft \textbf{Challenge 2: Handling sparse, skewed and correlated activities.}}
The time series activity data generated by each new user are multi-dimensional. As we show in Section \ref{sec:clustering}, such activity data are very sparse. For example, \textit{Chatters} are usually only active in the first four dimensions as described in Table \ref{tab:act}, while \textit{Sleepers} and \textit{Invitees} are inactive in most dimensions. Even \textit{All-star} users have a lot of 0's in certain dimensions. Besides the many 0's, the distributions of activity counts are highly skewed instead of uniform and many activities are correlated, like we discuss in Section \ref{sec:clustering}.1. 

{\flushleft \textbf{Challenge 3: Leveraging underlying user types.}}
As shown in our new user clustering analysis and highlighted in Figure \ref{fig:type} (b), our clustering of new users is highly indicative of user churn and should be leveraged for better churn prediction. However, as we only get access to initial several days instead of the whole two-week behaviors, user types are also unknown and should be jointly inferred with user churn. Therefore, how to design the proper model that can simultaneously learn the patterns for predicting user types and user churn poses a unique technical challenge that cannot be solved by existing approaches.

\subsection{Methods and Results}
We propose a series of solutions to treat the challenges listed above. Together they form our efficient churn prediction framework.
We also present comprehensive experimental evaluations for each proposed model component. Our experiments are done on an anonymous internal dataset of Snapchat, which includes 37M users and 697M bi-directional links.
The metrics we compute include accuracy, precision, and recall, which are commonly used for churn prediction and multi-class classification \cite{tsoumakas2006multi}.
The baselines we compare with are logistic regression and random forest, which are the standard and most widely practiced models for churn prediction and classification. 
We randomly split the new user data into training and testing sets with the ratio 8:2 for 10 times, and run all compared algorithms on the same splits to take the average performance for evaluation.
All experiments are run on a single machine with a 12-core 2.2GHz CPU and no GPU, although the runtimes of our neural network models can be largely improved on GPUs.

{\flushleft \textbf{Solution 1: Sequence-to-sequence learning with LSTM.}}
The intrinsic problem of user behavior understanding is sequence modeling. The goal is to convert sequences of arbitrary lengths with temporal dependences into a fixed-length vector for further usage. To this end, we propose to leverage the state-of-the-art sequence-to-sequence model, that is, LSTM  (Long-Short Term Memory) from the family of RNN (Recurrent Neural Networks) \cite{hochreiter1997long, mikolov2010recurrent}. Specifically, we apply a standard multi-layer LSTM to the multi-dimensional input $\mathbf{A}$. Each layer of the LSTM computes the following functions
\begin{align}
i_t&=\sigma(W_i\cdot[h_{t-1}, x_t]+b_i) \\\nonumber
f_t&=\sigma(W_f\cdot[h_{t-1}, x_t]+b_f) \\\nonumber
c_t&=f_t*c_{t-1}+i_t*\text{tanh}(W_c\cdot[h_{t-1}, x_t]+b_c) \\\nonumber
o_t&=\sigma(W_o\cdot[h_{t-1}, x_t]+b_o) \\\nonumber
h_t&=o_t*\text{tanh}(c_t)
\end{align}
where $t$ is the time step in terms of days, $h_t$ is the hidden state at time $t$, $c_t$ is the cell state at time $t$, $x_t$ is the hidden state of the previous layer at time $t$, with $x_t=a_{\cdot t}$ for the first layer, and $i_t$, $f_t$, $o_t$ are the input, forget and out gates, respectively. $\sigma$ is the sigmoid function $\sigma(x)=1/(1+e^{-x})$. Dropout is also applied to avoid overfitting. We use $\Theta_l$ to denote the set of parameters in all LSTM layers.

A linear projection with a sigmoid function is connected to the output of the last LSTM layer to produce user churn prediction as
\begin{align}
\hat{y} = \sigma(\mathbf{W}_c o_T+\mathbf{b}_c).
\label{eq:linear}
\end{align}
We use $\Theta_c$ to denote the parameters in this layer, \ie, $\mathbf{W}_c$ and $\mathbf{b}_c$.

Unlike standard methods for churn prediction such as logistic regression or random forest, LSTM is able to model user behavior data as time series and capture the evolvement of user activities through recognizing the intrinsic temporal dependencies. Furthermore, compared with standard time series models like HMM, LSTM is good at capturing both long term and short term dependences within sequences of variable lengths. When the lengths are short, LSTM acts similarly as basic RNN  \cite{mikolov2010recurrent}, but when more user behaviors become available, LSTM is expected to excel. 

\begin{figure*}[h!]
\centering
\vspace{-20pt}
\subfigure[Single LSTM]{
\includegraphics[width=0.235\textwidth]{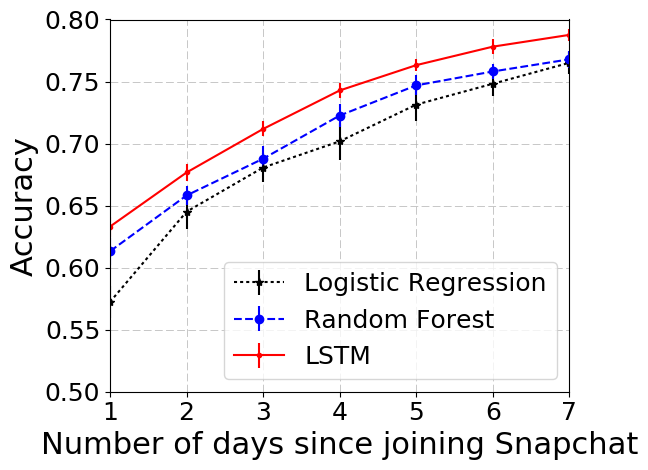}}
\subfigure[Activity embedding]{
\includegraphics[width=0.235\textwidth]{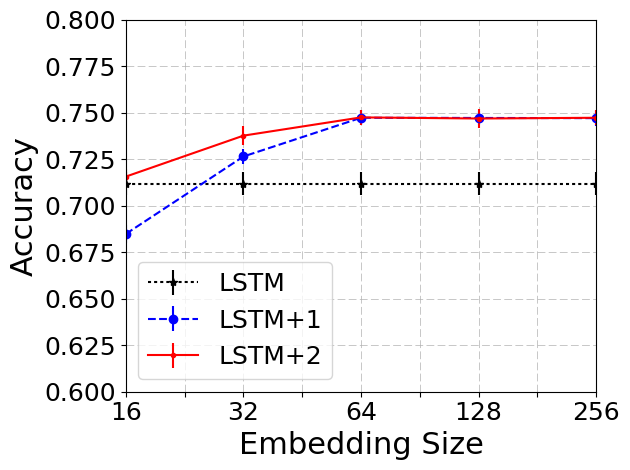}}
\subfigure[Parallel LSTMs]{
\includegraphics[width=0.235\textwidth]{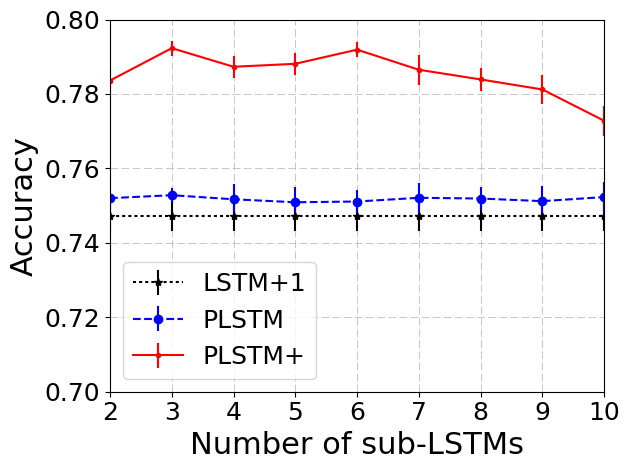}}
\subfigure[User type prediction]{
\includegraphics[width=0.26\textwidth]{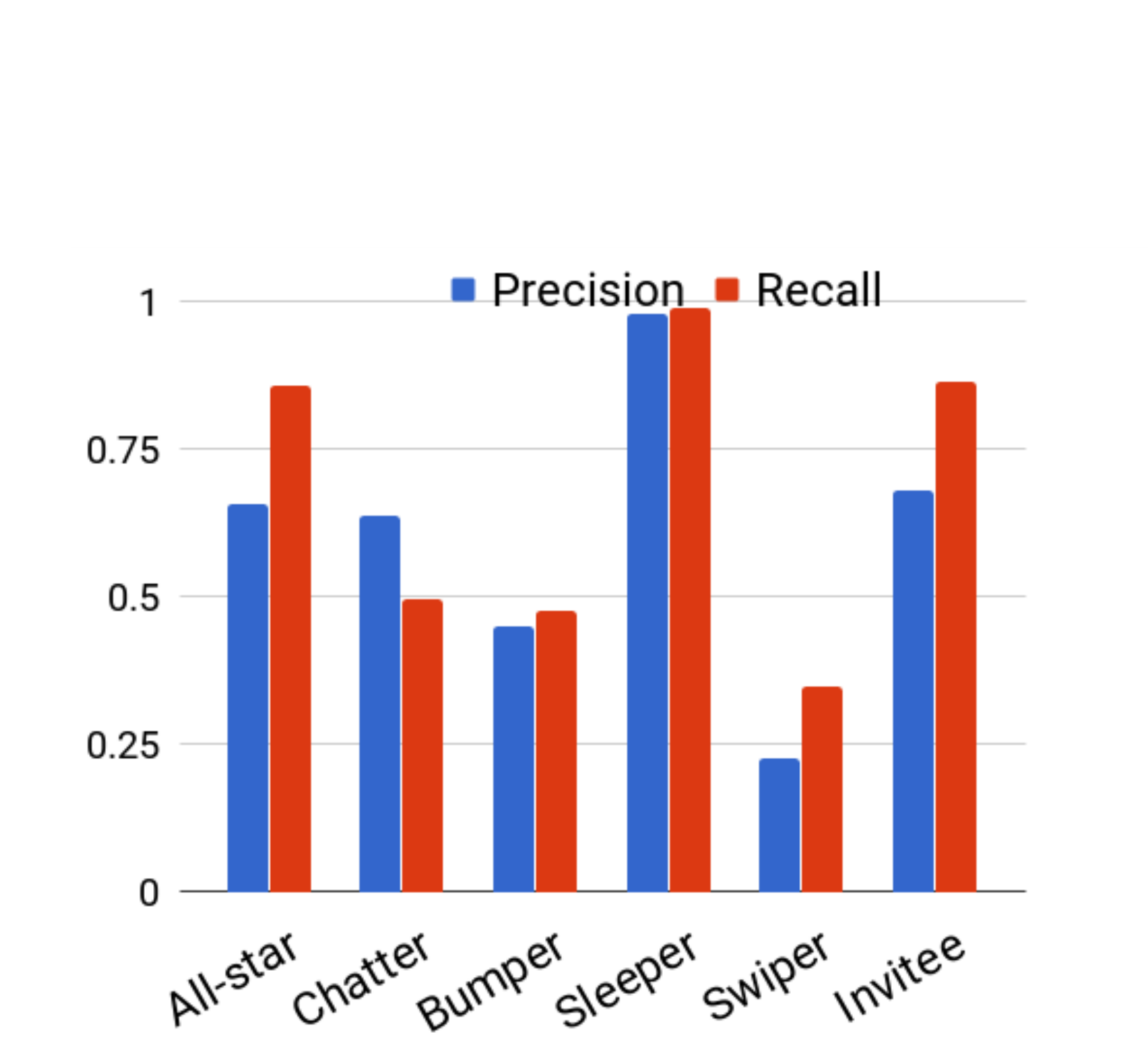}}
\caption{\textbf{Comprehensive experimental results on our churn prediction framework compared with various baseline methods.}}
\label{fig:exp}
\end{figure*}

Figure \ref{fig:exp} (a) shows the performances of compared models. 
The length of the output sequence of LSTM is empirically set to 64. 
In the experiments, we vary the amounts of user behavior data the models get access to and find that more days of behavior data generally lead to better prediction accuracy. We can also see that new users' initial activities in the first few days are more significant in improving the overall accuracy. A simple LSTM model can outperform all compared baselines with a substantial margin. The runtime of LSTM on CPU is within ten times of the runtimes of other baselines, and it can be significantly improved on GPUs.

{\flushleft \textbf{Solution 2: LSTM with activity embedding.}}
To deal with sparse, skewed and correlated activity data, we propose to add an activity embedding layer in front of the standard LSTM layer. Specifically, we connect a fully connected feedforward neural network to the original daily activity vectors, which converts users' sparse activity features of each day into distributional activity embeddings, while deeply exploring the skewness and correlations of multiple features through the linear combinations and non-linear transformations. 
Specifically, we have
\begin{align}
e_{\cdot t} = \psi^H(\ldots\psi^2(\psi^1(a_{\cdot t}))\ldots),
\end{align}
where
\begin{align}
\psi^h(e) = \text{ReLU}(W_e^h \text{Dropout}(e)+b^h_e).
\end{align}
$H$ is the number of hidden layers in the activity embedding network. $\Theta_e$ is the set of parameters in these $H$ layers.
With the activity embedding layers, we simply replace $\mathbf{A}$ by $\mathbf{E}$ for the input of the first LSTM layer, with the rest of the architectures unchanged.

Figure \ref{fig:exp} (b) shows the performances of LSTM with activity embedding of varying number of embedding layers and embedding sizes. 
The length of the output sequence of LSTM is kept as 64.
The overall performances are significantly improved with one single layer of fully connected non-linear embedding (\textit{LSTM+1}), while more layers (\eg, \textit{LSTM+2}) and larger embedding sizes tend to yield similar performances. The results are intuitive because a single embedding layer is usually sufficient to deal with the sparsity, skewness, and correlations of daily activity data. We do not observe significant model overfitting due to the dropout technique and the large size of our data compared with the number of model parameters.

{\flushleft \textbf{Solution 3: Parallel LSTMs with joint training.}}
To further improve our churn prediction, we pay attention to the underlying new user types. The idea is that, for users in the training set, we get their two-week behavior data, so besides computing their churn labels $y$ based on their second-week activities, we can also compute their user types $\mathbf{t}$ with our clustering framework. For users in the testing set, we can then compare the initial behaviors with those in the training set to guess their user types, and leverage the correlation between user types and churn for better churn prediction.

To implement this idea, we propose parallel LSTMs with joint training. Specifically, we assume there are $K$ user types. $K$ can be either chosen automatically by our clustering framework or set to specific values. Then we jointly train $K$ sub-LSTMs on the training set. Each sub-LSTM is good at modeling one type of users. We parallelize the $K$ sub-LSTMs and merge them through attention \cite{denil2012learning} to jointly infer hidden user types and user churn.

\begin{figure}[h!]
    \includegraphics[width=1\linewidth]{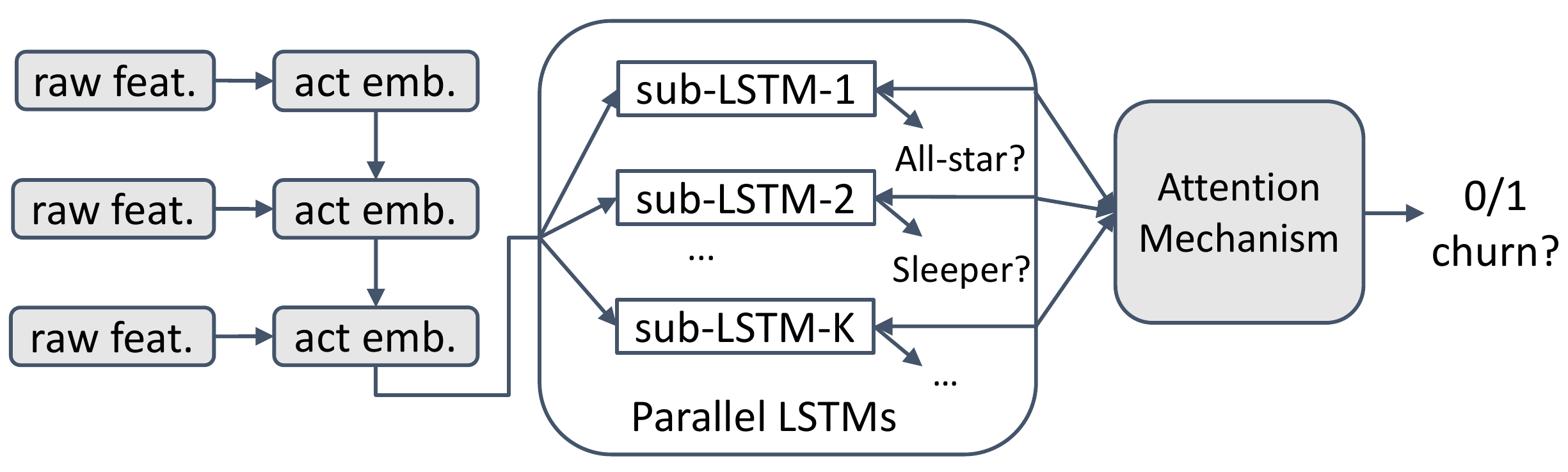}
    \caption{\small Parallel LSTMs with user type attention.}
    \label{fig:plstm}
\end{figure}

As shown in Figure \ref{fig:plstm}, for each user, the input of a sequence of activity embedding vectors $\mathbf{E}$ is put into $K$ sub-LSTMs in parallel to generate $K$ \textit{typed sequences}:
\begin{align}
\mathbf{s}_k = \text{LSTM}_k(\mathbf{E}).
\end{align}

To differentiate hidden user types and leverage this knowledge to improve churn prediction, we introduce an attention mechanism to generate user behavior embeddings by focusing on their latent types. A positive attention weight $w_k$ is placed on each user type to indicate the probability of the user to be of a particular type. We compute $w_k$ as a similarity of the corresponding typed sequence $\mathbf{s}_k$ and a global unique \textit{typing vector} $\mathbf{v}_t$, which is jointly learned during the training process. 
\begin{align}
w_k=\text{softmax}(\mathbf{v}_t^T \mathbf{s}_k).
\end{align}
Here \textsf{softmax} is taken to normalize the weights and is defined as $\text{softmax}(x_i)=\frac{\text{exp}(x_i)}{\sum_j \text{exp}(x_j)}$.
The user behavior embedding $\mathbf{u}$ is then computed as a sum of the typed sequences weighted by their importance weights:
\begin{align}
\mathbf{u}=\sum_{k=1}^K w_k \mathbf{s}_k.
\end{align}
The same linear projection with sigmoid function as in Eq.~\ref{eq:linear} is connected to $\mathbf{u}$ to predict user churn as binary classification.
\begin{align}
\hat{y} = \sigma(\mathbf{W}_c \mathbf{u}+\mathbf{b}_c).
\end{align}

To leverage user types for churn prediction, we jointly train a typing loss $l_t$ and a churn loss $l_c$.
For $l_t$, we firstly compute users' soft clustering labels $\mathbf{Q}$ as
\begin{align}
q_{ik} = \frac{(1+||\mathbf{f}_i-\mathbf{c}_k||^2)^{-1}}{\sum_j(1+||\mathbf{f}_i-\mathbf{c}_j||^2)^{-1}}.
\end{align}
$q_{ik}$ is a kernel function that measures the similarity between the feature $\mathbf{f}_i$ of user $u_i$ and the cluster center $\mathbf{c}_k$. It is computed as the probability of assigning $u_i$ to the $k$th type, under the assumption of Student's $t$-distribution with degree of freedom set to 1 \cite{maaten2008visualizing}.

We use $w_{ik}$ to denote the attention weight for user $u_i$ on type $t_k$. Thus, for each user $u_i$, we compute her typing loss as the cross entropy on $q_{i\cdot}$ and $w_{i\cdot}$. So we have
\begin{align}
l_t=-\sum_i \sum_k q_{ik}\log(w_{ik}).
\end{align}
For $l_c$, we simply compute the log loss for binary predictions as
\begin{align}
l_c=\sum -y_i \log\hat{y}_i - (1-y_i)\log(1-\hat{y}_i),
\end{align}
where $y_i$ is the binary ground-truth churn label and $\hat{y}_i$ is the predicted churn label for user $u_i$, respectively.

Subsequently, the overall objective function of our parallel LSTM with joint training is 
\begin{align}
l=l_c+\lambda l_t,
\end{align}
where $\lambda$ is a hyper-parameter controlling the trade-off between churn prediction and type prediction. We empirically set it to a small value like 0.1 in our experiments.

Figure \ref{fig:exp} (c) shows the performances of parallel LSTMs with and without joint training (\textit{PLSTM}+ vs.~\textit{PLSTM}). The only difference between the two frameworks is that \textit{PLSTM} is not trained with the correct user types produced by our clustering framework. In the experiments, we vary the number of clusters and sub-LSTMs and find that joint training is always significantly helpful. The performance of parallel LSTMs with joint training peaks with 3 or 6 sub-LSTMs. While the number 3 may accidentally align with some trivial clusters, the number 6 actually aligns with the six interpretable cohorts automatically chosen by our clustering framework, which illustrates the coherence of our two frameworks and further supports the sanity of splitting the new users into six types.

Besides churn prediction, Figure \ref{fig:exp} (d) shows that we can also predict what type a new user is by looking at her initial behaviors rather than two-week data, with different precisions and recalls. Our algorithm is good at capturing \textit{All-star}, \textit{Sleeper} and \textit{Invitee}, due to their distinct behavior patterns. \textit{Swiper} and \textit{Bumper} are harder to predict because their activity patterns are less regular. Nonetheless, such fast-response churn predictions with insights into user types can directly enable many actionable production decisions such as fast retention and targeted promotion. 

\section{Related Work}
\label{sec:related}

\subsection{User Modeling}
The problem of user modeling on the internet has been studied since the early 80's \cite{rich1979user}, with the intuitive consideration of stereotypes that can group and characterize individual users. However, during that time, with the scarce data, a lot of knowledge has to be manually collected or blindly inferred. Such labor and uncertainty make the models hard to capture stereotypes accurately.
While the lack of large datasets and labeled data have hindered deep user modeling for decades \cite{yin2015dynamic, webb2001machine}, the recent rapid growth of online systems ranging from search engine and social media to vertical recommenders have been collecting vast amounts of data and generating tons of new problems, which has enabled the resurgence of machine learning in user modeling. 

Nowadays, there has been a trend in fusing personalization into various tasks, including search \cite{speretta2005personalized, guha2015user}, rating prediction \cite{tang2015user, golbeck2006filmtrust}, news recommendation \cite{abel2013twitter, liu2010personalized, yang2017bi}, place recommendation \cite{zhang2013igslr, tang2013exploiting, li2016point, yang2017bridging}, to name a few. Most of them design machine learning algorithms to capture users' latent interests and mutual influences. However, while boosting the overall performance for particular tasks, such algorithms work more like black boxes, without yielding interpretable insight into the user behaviors.

Instead of building the model based on vague assumptions of user behavior patterns, in this work, we systematically analyze users' activity data and strive to come up with a general framework that can find interpretable user clusters. For any real-world online system as we consider, such interpretation can lead to both better prediction models and more personalized services.

\subsection{Churn Prediction}
It has been argued for decades that acquiring new users is often more expensive than keeping the old ones \cite{daly2002pricing, gillen2005winning}. Surprisingly, however, user retention and its core component, churn prediction, have received much less attention from the research community.
Only a few papers can be found discussing user churn, by modeling it as an evolutionary process \cite{au2003novel} or based on network influence \cite{kawale2009churn}.
While significant research efforts have been invested on brilliant ads to acquire new users, when dealing with old users, the most common practice is to simply plug in off-the-shelf logistic regression or random forest model\footnote{http://blog.yhat.com/posts/predicting-customer-churn-with-sklearn.html}\footnote{https://www.dataiku.com/learn/guide/tutorials/churn-prediction.html}. 

To the best of our knowledge, this is the first effort in the research community to seriously stress the emergence and challenge of churn prediction for online platforms. 
We are also the first to shift the focus of churn prediction from black box algorithms to deep understanding while maintaining high performance and scalability.

\subsection{Network Analysis}
Recent algorithms on network analysis are mostly related to the technique of network embedding \cite{perozzi2014deepwalk, grover2016node2vec, tang2015line, cao2015grarep, yang2017cone, yang2015network, niepert2016learning, kipf2016semi, defferrard2016convolutional, yang2018did}. They are efficient in capturing the high-order similarities of nodes regarding both structural distance and random-walk distance in large-scale networks. However, the latent representations of embedding algorithms are hard to interpret, as they do not explicitly capture the particular essential network components, such as hubs, cliques, and isolated nodes. Moreover, they are usually static and do not capture the evolution or dynamics of the networks.

On the other hand, traditional network analysis mostly focuses on the statistical and evolutionary patterns of networks \cite{epasto2015ego, leskovec2005graphs}, which provides more insights into the network structures and dynamics. For example, the early work on Bowtie networks \cite{arasu2002pagerank} offers key insight into the overall structure of the web; more recent works like \cite{danescu2013no, lo2016understanding, mcauley2013amateurs} help people understand the formation and evolution of online communities; the famous Facebook paper analyzes the romantic partnerships via the dispersion of social ties \cite{backstrom2014romantic} and so on. Such models, while providing exciting analysis of networks, do not help improve general downstream applications.

In this work, we combine traditional graph analysis techniques such as ego-network construction, degree and density analysis, as well as network core modeling, together with advanced neural network models, to coherently achieve high performance and interpretability on user clustering and churn prediction.

\subsection{Deep Learning}
While much of the current research on deep learning is focused on image processing, we mainly review deep learning models on sequential data, because we model user activities as multi-dimensional time series. Current research on deep learning has agreed that RNN \cite{mikolov2010recurrent} is the best model for sequential data. Its network design with loops allows information to persist and has been widely used in tasks like sentiment classification \cite{tang2015document}, image captioning \cite{mao2014deep} and language translation \cite{cho2014learning}, as a great substitute to traditional models like HMM. Among many RNN models, LSTM \cite{hochreiter1997long}, which specifically deals with long-term dependencies, is often the most popular choice. Variants of LSTM have also been developed, such as GRU (Gated Recurrent Unit) \cite{cho2014learning}, bi-directional LSTM \cite{schuster1997bidirectional}, tree LSTM \cite{tai2015improved}, latent LSTM \cite{zaheer2017latent} and parallel LSTM \cite{bouaziz2016parallel}. They basically tweak on the design of LSTM neural networks to achieve better task performance, but the results are still less interpretable.

In this work, we leverage the power of LSTM, but also care about its interpretability. Rather than using neural networks as black boxes, we integrate it with an interpretable clustering pipeline and leverage the hidden correlations among user types and user churn with an attention mechanism. 
Attribute embedding is also added to make the model work with sparse noisy user behavior data.

\section{Conclusions}
\label{sec:con}
In this paper, we conduct in-depth analysis of Snapchat's new user behavior data and develop \textit{ClusChurn}, a coherent and robust framework for interpretable new user clustering and churn prediction. Our study provides valuable insights into new user types, based on their daily activities and network structures, and our model can accurately predict user churn jointly with user types by taking limited data of new users' initial behaviors after joining Snapchat on large scales. While this paper focuses on the Snapchat data as a comprehensive example, the techniques developed here can be readily leveraged for other online platforms, where users interact with the platform functions as well as other users. 

We deployed \textit{ClusChurn} in Snap Inc.~to deliver real-time data analysis and prediction results to benefit multiple productions including user modeling, growth, retention, and so on. 
Future works include but are not limited to the study on user ego-network heterogeneity, where we hope to understand how different types of users connect with each other, as well as the modeling of user evolution patterns, where we aim to study how users evolve among different types and how such evolvements influence their activeness and churn rates.

\bibliographystyle{ACM-Reference-Format}
\bibliography{carlyang} 
\end{document}